\documentclass[aps,pra,twocolumn,superscriptaddress,10pt]{revtex4-1} 

\usepackage{amsmath,amssymb}
\usepackage{mathrsfs}
\usepackage{epstopdf}
\usepackage{graphicx}
\usepackage{bm,color,bbm}
\usepackage{natbib}
\usepackage{hyperref}
\usepackage{caption}
\usepackage{subcaption}

\newcommand{\ket}[1]{ |{#1} \rangle}

\newcommand{\Ket}[1]{\left|{#1}\right\rangle}

\newcommand{\sq}[1]{\left[ {#1} \right]}
\newcommand{\cu}[1]{\left\{ {#1} \right\}}
\newcommand{\ro}[1]{\left( {#1} \right)}

\newcommand{\ex}[1]{\langle{#1}\rangle }

\newcommand{\pauli}[2]{\sigma_{#2}^{(#1)}}

\newcommand{\spin}[1]{\hat{\Sigma}_{#1}}
\newcommand{\spinb}[1]{\bar{\Sigma}_{#1}}
\newcommand{\spint}[1]{\tilde{\Sigma}_{#1}}

\newcommand{\Tr}[1]{\operatorname{Tr}\ro{#1}}
\newcommand{\abs}[1]{\left| {#1} \right|}

\begin{document}

\preprint{}

\title{Central Moment Analogues to Linear Optically Generated Cluster States}


\author{Christopher C. Tison}
\email{ctison@fau.edu}
\affiliation{Air Force Research Laboratory, Information Directorate, Rome, New York, 13441, USA}
\affiliation{Department of Physics, Florida Atlantic University, Boca Raton, Florida, 33431, USA}
\affiliation{Quanterion Solutions Incorporated, Utica, New York, 13502, USA}
\author{James Schneeloch}
\email{jfizzix@gmail.com}
\affiliation{Air Force Research Laboratory, Information Directorate, Rome, New York, 13441, USA}
\author{Paul M. Alsing}
\email{paul.alsing@us.af.mil}
\affiliation{Air Force Research Laboratory, Information Directorate, Rome, New York, 13441, USA}


\date{\today}

\begin{abstract}
Two-mode squeezed states in the limit of small squeezing, Hong-Ou-Mandel interference and post selection on coincidence counts are some of the staples of linear quantum optics. 
We show that by using classical expectations on central moments of intensities, we can remove the requirement of small squeezing necessary for high fidelity coincidence detection.
Utilizing existing techniques to probabilistically generate a cluster state, we construct a statistical analogue with deterministic generation at the cost of losing the ability to feed forward and requiring statistical averaging. 
\end{abstract}



\maketitle

Cluster states are a paradigm of quantum computing by which conditional measurements are utilized to deterministically enact quantum logic on an entangled state \cite{clusterstate, feed_forward}. Nonlinear interaction in the form of a controlled phase gate is the canonical form of generating such an entangled state. Sadly, the nonlinearity necessary to deterministically enact a controlled phase gate on single photons has proven elusive. Accordingly, the toolbox for generating entangled single photon states hinges on linear optics and post selection \cite{pan_4_mode}. Post selection acts as an effective nonlinearity at the cost of probabilistic success.  Much work has been put into preventing this probabilistic approach from becoming intractable for arbitrarily large clusters \cite{3photonGHZ}. On the other end of the spectrum, single photon generation is inherently stochastic and in the case of spontaneous parametric down conversion, it has an upper limit where multi-photon generation becomes non-negligible and contaminates the post selection conditions. This is mitigable through off-line preparation of photon pairs and photon number resolved heralding, now at the cost of needing to store photons \cite{silberhorn}.

We address the problem of single photon generation by mapping single photon statistics to intensity fluctuations on bright beams \cite{braunstein_review}. This approach is distinct from previous efforts, which also use bright beams, but focus on the quadrature uncertainties of squeezed fields \cite{furusawa_large, menicucci_cv_cluster}. Both approaches with respect to single photons trade probability of occurrence with certainty of outcome. In particular, we change from photon pair generation via spontaneous parametric down-conversion to two-mode squeezed states and map coincidences to cross central moment expectations. Using this mapping we find analogous statistics to Hong-Ou Mandel interference \cite{HOM87}, Bell states \cite{kwiat95}, GHZ states \cite{PAN8GHZ}, and finally cluster states \cite{feed_forward}. Along the way we find loss and amplification can be used to manipulate the central moments. Lastly, we find that the nature of using only linear optics invokes an exponential increase in the uncertainty of our measures as more modes are added to the state. Conversely, we do not have the up front cost of probabilistically generating our photons and do not require single photon detectors.

The simplest example of our mapping is in observing an analogy to Hong-Ou-Mandel interference.  We generalize single photon pairs to the small squeezing regime of a two-mode squeezer described by the unitary operator \cite{agarwal}
\begin{align}
\hat{S}_{ab}(r,\phi)=\exp{\sq{ r e^{i\phi} \hat{a}^\dagger \hat{b}^\dagger- r e^{-i\phi} \hat{a}\hat{b} }},
\end{align}
for positive real values $r$ and $\phi$ which determine the amount of squeezing and relative phase of the generated modes $a$ and $b$. The two-mode squeezer acts on the vacuum by
\begin{subequations}	
\begin{align}
\hat{S}_{ab}(r,\phi)\ket{\text{vac}} 
      &= 
	\sum_{n=0}^\infty \frac{\left[\tanh(r) e^{i\phi} \right]^n}{\cosh(r)} \ket{n}_a\ket{n}_b \label{eq:tmss} \\
      &\approx
        \ket{\text{vac}} + r e^{i\phi} \ket{1}_a\ket{1}_b, \quad r\ll 1.\label{eq:spstate}
\end{align}
\end{subequations}
Normalization is omitted in \eqref{eq:spstate} to show the small $r$ limit of two-mode squeezing is approximated by photon pair generation. Measuring the Pearson correlation of their intensities $\hat{n}_i=\hat{a}_i^\dag \hat{a}_i$, shows that the two-mode squeezed vacuum exhibits perfect correlation. To see this, we note their central moment intensity operators are defined by $\bar{n}_i=\hat{n}_i-\ex{\hat{n}_i}$. The modes see equal average photon number $\ex{\hat{n}_a}=\ex{\hat{n}_b}=\sinh^2(r)$ and their variances are equal to their covariance as
\begin{align}
	\ex{\bar{n}_a\bar{n}_a} = \ex{\bar{n}_b\bar{n}_b} = \ex{\bar{n}_a\bar{n}_b} = \cosh^2(r)\sinh^2(r),
\end{align}
leaving the Pearson correlation to be one
\begin{align}
	\label{eq:pearson}
	\rho_{ab}=\ex{\bar{n}_a\bar{n}_b}/\sqrt{\ex{\bar{n}_a\bar{n}_a}\ex{\bar{n}_b\bar{n}_b}}=1,
\end{align}
independent of squeezing, $r$. Pearson correlation is a measure of linear correlation between two random variables. It is bounded between plus and minus one corresponding to correlation and anti-correlation, respectively. Therefore, \eqref{eq:pearson} shows perfect linear correlation is present in the two mode squeezed state.

We can remove all Pearson correlation by using the Hadamard operation $\hat{H}_{ab}$. The Hadamard operator transforms the modes $a$ and $b$ as
\begin{align}
	\hat{a} \rightarrow (\hat{a} + \hat{b})/\sqrt{2} = \hat{a}^\prime, \quad
	\hat{b} \rightarrow (\hat{a} - \hat{b})/\sqrt{2} = \hat{b}^\prime, 
\end{align}
and describes the operation of a $50/50$ beamsplitter on two spatial modes. 
Measuring the state \eqref{eq:tmss} in this new basis shows there are no correlations between $a^\prime$ and $b^\prime$ as their covariance $\ex{\bar{n}_{a^\prime}\bar{n}_{b^\prime}}=0$. The Hadamard operation transforms between maximal and zero intensity correlation on the two-mode squeezed state \cite{macroHOM}. This is in analogy with Hong-Ou-Mandel interference which sees the Hadamard operation transform two photons from perfect coincidence to zero coincidence.


Intensity detection can also be used to measure Bell state statistics. In the limit of small squeezing, the $\ro{\ket{HH} + \exp(i\phi)\ket{VV}}/\sqrt{2}$ state can be approximated by the application of two squeezing operations
\begin{subequations}
\begin{align}
	\label{eq:bell}
	\ket{\psi_\text{Bell}} =&
	\hat{S}_{a_hb_h}(r,0)\hat{S}_{a_vb_v}(r,\phi)\ket{\text{vac}}  \\
	\label{eq:approxbell}
	\approx &
	 \ket{\text{vac}}+ r \ro{\hat{a}^\dag_h\hat{b}^\dag_h + e^{i\phi} \hat{a}^\dag_v\hat{b}^\dag_v }\ket{\text{vac}},
\end{align}
\end{subequations}
where in \eqref{eq:approxbell} $r\ll 1$ and we ignored normalization. The subscripts $h$ and $v$ are merely markers to denote the additional modes; therefore all four Bell states can be generated in a similar way. In the case where these subscripts denote polarizations, one can use the Stokes parameters \cite{gabay, pfister_stokes}
\begin{subequations} 
\begin{align}
\spin{a_1} &= \hat{a}^\dag_h \hat{a}_h - \hat{a}^\dag_v \hat{a}_v, \\
\hat{\Sigma}_{a_2} &= \hat{a}^\dag_h \hat{a}_v +\hat{a}^\dag_v \hat{a}_h, \\
\hat{\Sigma}_{a_3} &=i \ro{ \hat{a}^\dag_v \hat{a}_h-\hat{a}^\dag_h \hat{a}_v },
\end{align}
\end{subequations}
and total photon number $\hat{\Sigma}_{a_0} = \hat{a}^\dag_h \hat{a}_h + \hat{a}^\dag_v \hat{a}_v$ to perform polarization measurements. These operators correspond to photon number resolution in the low photon flux regime, and photocurrent measurement of p-i-n diode detectors in the high flux regime \cite{pin}. Using these operators, we can measure the two-mode squeezed Bell state (TMS-Bell) \eqref{eq:bell} and find the $a$ mode is not polarized
%
\begin{align}
 \ex{\spin{a_0}} = 2\sinh^2(r), \quad  \ex{\spin{a_i}} = 0, \quad i=\cu{1,2,3},
\end{align}
where we make no assumption on the magnitude of $r$ and a similar relation holds for the $b$ mode. For concreteness we restrict to the case where $\phi=0$. The nonzero joint expectations of Stokes parameters (and total intensity) between modes $a$ and $b$ are
\begin{subequations}\label{eq:bellstat}
\begin{align}
\ex{\spin{a_0}\spin{b_0}} &= 2\sinh^2(r)\cosh^2(r) + 4\sinh^4(r), \label{eq:zzeq}\\
\ex{\spin{a_1}\spin{b_1}} &= 2\sinh^2(r)\cosh^2(r), \\
\ex{\spin{a_2}\spin{b_2}} &= 2\sinh^2(r)\cosh^2(r), \\
\ex{\spin{a_3}\spin{b_3}} &= -2\sinh^2(r)\cosh^2(r).
\end{align}
\end{subequations}
Here, we want to explicitly mention all cross terms ($\ex{\spin{a_i}\spin{b_j}}=0, i\ne j$) are zero. Using the convention of normalizing by the joint total photon number \eqref{eq:zzeq}, it has been shown that this state can violate a CHSH inequality for small values of $r$ but approaches the classical value of $2\sqrt{2}/3$ as $r$ is increased \cite{reid}. It should be clear that this is due to the contribution of $4\sinh^4(r)$ which is only present in the joint total photon number measurement \eqref{eq:zzeq}. Following similar footsteps to the single two-mode squeezed state, 
we propose the measurement of central moment statistics. Accordingly, 
\begin{align}
\ex{\spinb{a_0}\spinb{b_0}}\equiv\ex{\spin{a_0}\spin{b_0}}-\ex{\spin{a_0}}\ex{\spin{b_0}},
\end{align}
and the TMS-Bell state, \eqref{eq:bell} has balanced (all equal) statistics for central moments. Additionally, the variances $\ex{\spinb{a_i}\spinb{a_i}}=2\sinh^2(r)\cosh^2(r)$ (similarly for $b$ modes) are the same magnitude, revealing perfect Pearson correlation for each of these measurements. We have to note, the derivation of the CHSH inequality does not work with central moment statistics. Therefore, we are not claiming a CHSH inequality violation despite the TMS-Bell state producing Bell like statistics.

As further proof that our state cannot violate a CHSH inequality, we find that in the presence of linear loss or misbalanced squeezing, the central moment statistics of \eqref{eq:bell} can always be rebalanced through the introduction of more loss or amplification. This comes at the cost of reducing Pearson correlation. Nevertheless, it is in stark contrast to quadrature based continuous variable entanglement which is sensitive to loss and single photon states which cannot be amplified.

Linear loss can be modeled as a beamsplitter with an unmeasured Bosonic mode $\hat{\gamma}_l$ which transforms annihilation operators as
\begin{align}
\hat{a}\rightarrow \cos(\theta) \hat{a} + i \sin(\theta) \hat{\gamma}_l
\end{align}
and $0\leq \theta \leq \pi/2$ indicates zero to total loss, respectively. In the case where there are multiple sources of loss, we assume their unmeasured modes are different such that they commute as $[\hat{\gamma}_{l_i},\hat{\gamma}^\dag_{l_j}]=0$.

Conversely, a two-mode squeezer acts as an ideal linear amplifier. It adds vacuum fluctuations to the mode $\hat{a}$ as
\begin{align}
\hat{a}\rightarrow \cosh(g) \hat{a} + \sinh(g) \hat{\gamma}_g^\dag
\end{align}
where, $g\ge 0$ is the gain and $\hat{\gamma}_g$ is added noise (which is uncorrelated as in the linear loss case.)

In an attempt to better represent the statistics we adopt a correspondence between the central moment spin measures and Pauli spin matrices as 
\begin{align}
\ex{\spinb{a_i}\spinb{b_j}} \rightarrow \ex{\spinb{a_i}\spinb{b_j}} \sigma_i\otimes\sigma_j
\end{align}
(where $\sigma_0$ is the identity and $\sigma_{\cu{1,2,3}}\leftrightarrow\sigma_{\cu{z,x,y}}$).  Accordingly, we represent the unequally squeezed state $\hat{S}_{a_hb_h}(r_1,0)\hat{S}_{a_vb_v}(r_2,0)\ket{\text{vac}}$ as
\begin{subequations}\label{eq:density}
\begin{align} 
\rho_\text{Bell} &= \sum_{\cu{i,j}=0}^3 \ex{\spinb{a_i}\spinb{b_j}} \sigma_i\otimes\sigma_j \\
&= \sq{
\begin{matrix}
\eta & 0 & 0 & \mu \\
0 & 0 & 0 & 0 \\
0 & 0 & 0 & 0 \\
\mu & 0 & 0 & \nu \\
\end{matrix}},\text{ } \left\{
\begin{matrix}
\eta =& \sinh^2(r_1)\cosh^2(r_1) \\
\mu =& \sinh(r_1)\cosh(r_1)\\
 \phantom{=}&  \times\sinh(r_2)\cosh(r_2) \\
\nu =& \sinh^2(r_2)\cosh^2(r_2) \\
\end{matrix}
\right.
\end{align}
\end{subequations}
revealing the suggestive Bell statistics but we must note the usage of central moments prevents this from being a valid density matrix. In the case of $r_1>r_2$, the introduction of the same loss to both $a_h$ and $b_h$ takes the coefficient $\eta \rightarrow \cos^4(\theta) \eta$ and the coefficient $\mu\rightarrow \cos^2(\theta) \mu$. Therefore, we can pick the loss to be such that $\cos^2(\theta)\sinh(r_1)\cosh(r_1)=\sinh(r_2)\cosh(r_2)$, rebalancing the statistics.

In a similar approach, one can find that applying equal and independent gain to the $a_v$ and $b_v$ modes transforms $\mu\rightarrow \cosh^2(g) \mu$ and $\nu\rightarrow \cosh^4(g) \nu$. In this manner, a choice of gain where $\sinh(r_1)\cosh(r_1)=\cosh^2(g)\sinh(r_2)\cosh(r_2)$ is appropriate.

As mentioned earlier, these actions harm the Pearson correlations of the measurements. In fact, the variances no longer equal each other and in order to characterize the state; the sensible thing to do is normalize by the trace such that $\Tr{\rho_\text{Bell}}=1$ or equivalently divide through by $\Tr{\sigma_0\otimes\sigma_0}\ex{\spinb{a_0}\spinb{b_0}}$. We must keep in mind the variances are indicators of the amount of time necessary to bound the statistics as ultimately, one needs to be capable of discriminating nonzero measures from ones that are zero.

Moving forward, one way to generate a four photon two-mode squeezed GHZ (TMS-GHZ) state is to generate two TMS-Bell states and use a polarizing beam splitter to mix their information. A polarizing beam splitter acts as a perfect transmitter for one polarization and a perfect reflector for the orthogonal one giving
\begin{align}
	\cu{\hat{a}_h,\hat{b}_h}\rightarrow\cu{\hat{a}_h,\hat{b}_h}, \quad 
	\cu{\hat{a}_v,\hat{b}_v}\rightarrow\cu{\hat{b}_v,\hat{a}_v}, \nonumber
\end{align}
where for simplicity, we ignore the phase gained upon reflection. We can incorporate the action of a polarizing beamsplitter (PBS) into the labeling of the squeezers giving
\begin{align} \label{eq:GHZ}
\ket{\psi_\text{GHZ}}=\hat{S}_{a_h b_h}\hat{S}_{a_v d_v}\hat{S}_{c_h d_h}\hat{S}_{c_v b_v}\ket{\text{vac}}
\end{align}
with the recognition that $(r,0)$ are implicit arguments to the squeezing operators and a swap between $b_v$ and $d_v$ was applied to the initial TMS-Bell states on modes $a$, $b$ and $c$, $d$. We find the measurement outcomes to be
\begin{align} 
\rho_\text{GHZ} &= \sum_{\cu{i,j,k,l}=0}^3 \ex{\spinb{a_i}\spinb{b_j}\spinb{c_k}\spinb{d_l}} \sigma_i\otimes\sigma_j\otimes\sigma_k\otimes\sigma_l \nonumber \\
&= \sq{
\begin{matrix}
\delta & 0 & \cdots & 0 & \delta \\
0 & 0 & \cdots & 0 & 0 \\
\vdots & \vdots & \ddots & \vdots & \vdots \\
0 & 0 & \cdots & 0 & 0 \\
\delta & 0 & \cdots & 0 & \delta \\
\end{matrix}},\text{ } \delta = \sinh^4(r)\cosh^4(r)
\end{align}
where $\rho_\text{GHZ}$ is of dimension $2^4\times2^4$ and normalization is given by dividing through with $\Tr{\rho_\text{GHZ}}=2\delta$. 

Generalizing the two mode case of Pearson correlations one can regularize the central moments via
\begin{align}
\spint{a_i} = \frac{\spinb{a_i}}{\sqrt{\ex{\spinb{a_i}\spinb{a_i}}}}.
\end{align}
We denote expectations across modes of these regularized operators as a measure of the co-fluctuation of the state. Just as in the TMS-Bell state \eqref{eq:bell}, the individual modes of the TMS-GHZ state have variances of $\ex{\spinb{a_i}\spinb{a_i}}=2\sqrt{\delta}$ while nonzero central moment measures of the TMS-GHZ state (i.e. $\ex{\spinb{a_0}\spinb{b_0}\spinb{c_0}\spinb{d_0}}$) all have magnitude $2\delta$. Therefore, the co-fluctuation of the TMS-GHZ state ends up being $\frac{1}{2}$. In contrast, the two TMS-Bell states before application of the PBS see a co-fluctuation of $1$. In the terminology of single photon statistics, we are seeing the well known result of $50\%$ success of post selecting coincidences after mixing on a polarizing beamsplitter \cite{PBS}. Waiting for a coincidence event to occur translates to a longer wait time necessary to bound the measurement statistics.


We would like to note that this procedure only works for combining an even number of modes. Additionally, if one was so inclined, they could find other ways to assign squeezing such that they reproduce the statistics. For instance, instead of swapping $b_v$ and $d_v$, one could have swapped $a_v$ and $c_v$. Finally, one could generalize to an $n$ mode TMS-GHZ state as
\begin{align}
\ket{\psi_\text{GHZ}}_n = \prod_{i=0}^{n-1} \hat{S}_{a^{(i)}_h a^{((i+1)\operatorname{mod}n)}_v}\ket{\text{vac}}
\end{align}
where $a^{(i)}$ are distinct modes, and $n= 2k$ for $k$ underlying TMS-Bell states. In the case of $k=1$, we recognize the $\ro{\ket{HV}+\ket{VH}}/\sqrt{2}$ Bell state statistics which for higher $k$ generalize accordingly and the magnitude of co-fluctuations drop off as $2^{1-k}$.

Combining Hadamard operations with polarizing beam splitters, one can extend beyond generation of GHZ states to cluster states. Cluster states are special states which allow a graphical representation to denote a vertex connectivity structure according to the eigenvalue relation 
\begin{align} \label{eq:cluster}
\sigma_2^{(i)}\bigotimes_{j\in \text{ngbh}(i)}\sigma_1^{(j)} \Ket{\psi} = \pm\Ket{\psi},
\end{align}
where $\text{ngbh}(i)$ denotes the set of vertices that share an edge with the vertex $i$ and there is a correspondence between the Pauli spin operators as $\sigma_1\equiv\sigma_z$ and $\sigma_2\equiv\sigma_x$. These eigenequations are known as stabilizer generators and they uniquely define the state  \cite{chuang}. Analogously, we define the TMS-cluster state to be one which obeys the TMS-stabilizer relation
\begin{align} \label{eq:approx_cluster}
\ex{\spinb{a_2^{(i)}}\prod_{j\in \text{ngbh}(i)}\spinb{a_1^{(j)}}\prod_{k \not\in \text{ngbh}(i)}\spinb{a_0^{(k)}} } = \beta\quad \forall i,
\end{align}
where $\not\in \text{ngbh}(i)$ is not inclusive of $i$ and $\abs{\beta}>0$. Note, we have gone from an eigenvalue relation to one of nonzero central moment expectations. This is because our central moment operators are not eigenoperators of the squeezed states. Additionally, we have to include the $\spinb{a_0}$ measures for all other vertices, even those which are not connected to the $i$th vertex. As the analogy goes, we are post selecting on coincidences and the measure $\spinb{a_0^{(i)}}$ is no longer simply the identity. 

\begin{figure}
\includegraphics[width=0.45\textwidth]{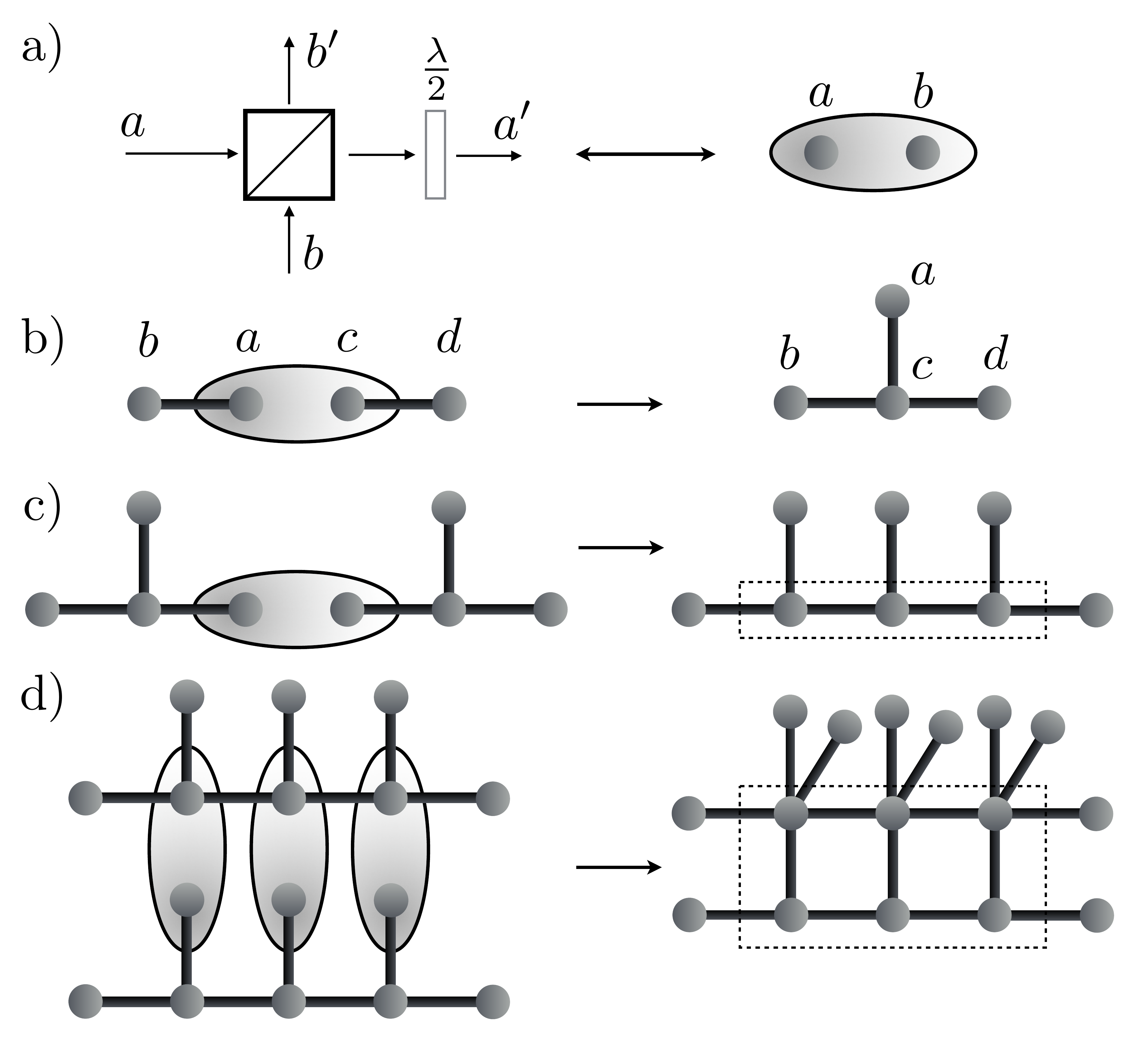}%
 \caption{a) Graphical representation for PBS gate of Bodiya and Duan \cite{duan}. b) Effect of PBS gate on $a$ and $c$ modes of two TMS-Bell states forming locally equivalent $4$ mode TMS-GHZ state with TMS-stabilizers in Eq \eqref{eq:tmsghz}. c) Combination of two $4$-mode TMS-clusters produces linear $3$-mode TMS-cluster indicated by a dashed box. d) Combining $2$ linear TMS-clusters to form a $2\times3$ planar cluster with $10$ leaves. In general, a $n\times m$ cluster by this approach will have $n(m + 2)$ leaves.}
 \label{fig:test}
\end{figure}

We form a $2D$ TMS-cluster state by following the footsteps of Bodiya and Duan \cite{duan}. All that is necessary are TMS-Bell states, Hadamard rotations, and polarizing beamsplitters. The action of a polarizing beamsplitter followed by measuring central moment expectation of the whole state means that components of the two beams incident on the PBS which go the same direction will be neglected. They do not share central moment expectation with the vacuum which occupies the other mode. Consequentially, the action of using a PBS on the state restricts one to only the correlated horizontal polarization statistics or correlated vertical polarization statistics. This manifests itself as an enforcement of the measures $\spinb{a_0}\spinb{b_0}$ and $\spinb{a_1}\spinb{b_1}$ being nonzero. 

Starting from the TMS-Bell state \eqref{eq:bell} we set the relative phase $\phi=0$, and a Hadamard operation on either the $a$ or $b$ mode changes the nonzero central moment expectations from
\begin{align}
&\ex{\spinb{a_0}\spinb{b_0}}=\ex{\spinb{a_1}\spinb{b_1}}=\ex{\spinb{a_2}\spinb{b_2}}=-\ex{\spinb{a_3}\spinb{b_3}} \nonumber \\
&\rightarrow \ex{\spinb{a_0}\spinb{b_0}}=\ex{\spinb{a_1}\spinb{b_2}}=\ex{\spinb{a_2}\spinb{b_1}}=\ex{\spinb{a_3}\spinb{b_3}}. \label{eq:tms-cluster}
\end{align}
The magnitude of $\ex{\spinb{a_0}\spinb{b_0}}=2\sqrt{\delta}$ is preserved under this operation and the latter represents a TMS-cluster state with nonzero TMS-stabilizer relations 
\begin{align}
\ex{\spinb{a_2}\spinb{b_1}}=\ex{\spinb{b_2}\spinb{a_1}}=2\sqrt{\delta}.
\end{align}

Starting with $2$ such TMS-cluster states on modes $a,b$ and $c,d$, a PBS on modes $a$ and $c$ transforms the TMS-stabilizer relation as
\begin{align}
&\ex{\spinb{a_2}\spinb{b_1}\spinb{c_0}\spinb{d_0}}=\ex{\spinb{b_2}\spinb{a_1}\spinb{c_0}\spinb{d_0}} \nonumber \\ 
&=\ex{\spinb{c_2}\spinb{d_1}\spinb{a_0}\spinb{b_0}}=\ex{\spinb{d_2}\spinb{c_1}\spinb{a_0}\spinb{b_0}} \\
\rightarrow&
\ex{\spinb{a_2}\spinb{b_1}\spinb{c_2}\spinb{d_1}}=\ex{\spinb{a_1}\spinb{b_0}\spinb{c_1}\spinb{d_0}} \nonumber \\
&=\ex{\spinb{a_0}\spinb{b_2}\spinb{c_1}\spinb{d_0}}=\ex{\spinb{a_0}\spinb{b_0}\spinb{c_1}\spinb{d_2}}, 
\end{align}
where the latter results are calculated analytically and in general do not have a simple relation to the preceding TMS-stabilizers. Additionally, the transformations show a mimicry of the involution property of Pauli matrices ($\sigma_i \sigma_i = \sigma_0$). For instance, the outcome $\ex{\spinb{a_0}\spinb{b_2}\spinb{c_1}\spinb{d_0}}$ is similar to a product of the stabilizer 
$\pauli{a}{1}\pauli{b}{2}$ ($\equiv  \sigma_1 \otimes \sigma_2$)
and the post selection on the PBS subspace of $\pauli{a}{1}\pauli{c}{1}$ such that
\begin{align}
&\ro{\pauli{a}{1}\pauli{b}{0}\pauli{c}{1}\pauli{d}{0}} 
\cdot \ro{\pauli{a}{1}\pauli{b}{2}\pauli{c}{0}\pauli{d}{0}}
= \pauli{a}{0}\pauli{b}{2}\pauli{c}{1}\pauli{d}{0}.
\end{align}
Following with a Hadamard operation on the $a$ mode we graphically represent this operation in Fig. \ref{fig:test}(a). In the syntax of Bodiya and Duan \cite{duan} we denote the usage of a PBS followed by Hadamard on one mode as a PBS operation. The resulting TMS-stabilizer is a star cluster with center $c$ as indicated in figure \ref{fig:test}b and TMS-stabilizers
\begin{align}
&\ex{\spinb{a_2}\spinb{c_1}\spinb{b_0}\spinb{d_0}}
=\ex{\spinb{b_2}\spinb{c_1}\spinb{b_0}\spinb{d_0}} \nonumber \\
&=\ex{\spinb{c_2}\spinb{a_1}\spinb{b_1}\spinb{d_1}}
=\ex{\spinb{d_2}\spinb{c_1}\spinb{a_0}\spinb{b_0}}.\label{eq:tmsghz}
\end{align}
The graphical procedure gives an effective means to see how one could continue to combine $4$ mode TMS-star clusters to form a linear TMS-cluster with additional vertices which only contain one connection. These vertices are known as leaves in the graph community. Using the leaves, one could then combine several linear clusters to form a $2$D cluster, combine many $2$D clusters to form a $3$D cluster and so on.

With respect to noise, the effect on co-fluctuation due to Bodiya and Duan's procedure falls in line with the results we have presented thus far. In particular, each usage of a PBS operation sees a drop in co-fluctuation by a factor of $2$.

Although the co-fluctuation decreases exponentially with the size of the state, the central moment expectations can grow. For instance, starting with $k$ Bell states, there are $n=2k$ spatial modes and the magnitude of central moment expectation is 
\begin{align}
\ex{\prod_{i=1}^n\spinb{a^{(i)}_0}}=\sq{\sqrt{2}\cosh\ro{r}\sinh\ro{r}}^n.
\end{align} 
If a loss of $0<\gamma\le1$ was introduced per spatial mode, the expectation would drop. For scalability, we require
\begin{align}
\sq{\gamma \sqrt{2}\cosh\ro{r}\sinh\ro{r}}^n\ge1\implies\sinh\ro{2r}\ge\frac{\sqrt{2}}{\gamma},
\end{align}
such that the nonzero measures of the state will be distinguishable from those that are. As an example, a modest value of $r=2.3$ (which correlates to an average of $\sim 49$ photons per spatial mode) can tolerate a modal loss of $\gamma=0.028\approx-15.5$ dB and independent of the size of the state, the central moment expectation would be nonzero. We can consume the loss tolerance to build a TMS-cluster via PBS operations. As mentioned previously, each PBS operation is equivalent to roughly $-3$ dB loss therefore, by applying it to each spatial mode $3$ times, one can generate a $2$D TMS-cluster and still have a loss budget of $-6.5$ dB per mode.

In order for single photon cluster states to deterministically perform computation, they require the capability of altering upcoming measurement settings dependent on previous measurement outcomes. This is known as feed forward. The decision making protocol relies on the binary nature of the measurements (a single photon will only trigger one detector) and modulo $2$ addition. Both the reliance on expectation values instead of eigenvalue relations and real valued intensity measures prevent an extension of these squeezed states to the feed forward protocol. Instead, we have to invoke another factor of $2$ hit in efficiency per spatial mode by post selecting on the measurement projections which do not require feed forward; dropping the loss budget of our $r=2.3$ example down to $-3.5$ dB.

We believe that this approach will allow one to get in the order of $20$ to $50$ modes before the statistics become too noisy to bound in a reasonable time. Additionally, the reliability of state generation and connections to the recently proposed measurement based classical computing protocol \cite{MBCC} may lead to a non-trivial experiment which leverages the power of these quantum type statistics to perform a calculation which is not efficiently computable by a classical computer.

In conclusion, we have explored the route of using central moment statistics on number correlated classical intensity states in lieu of post-selected coincidences on photon pair generated states. We show by using only linear optics to generate the state, central moment statistics are capable of exhibiting cluster state statistics that are non-vanishing for arbitrarily large states and finite loss. The tradeoff in using linear optics to build these TMS-cluster state statistics is an exponential increase in uncertainty of the measurement as a function of the size of the state. This approach is in contrast to post-selected single photons and linear optics which sees exponential wait times in coincidence detection both due to post-selection and the stochastic nature of their generation process. 

\begin{acknowledgments}
CCT, JS and PMA would like to acknowledge support of this work from 
Office of the Secretary of Defense ARAP QSEP program. 
JS would like to acknowledge support from the National Science Foundation.
CCT would also like to thank Gregory A. Howland and Michael L. Fanto for helpful discussions.
Any opinions, findings and conclusions or recommendations
expressed in this material are those of the author(s) and do not
necessarily reflect the views of Air Force Research Laboratory.
\end{acknowledgments}

\bibliographystyle{apsrev4-1}
\bibliography{CV_central_moments}

\end{document}